\documentclass[12pt]{article}
\usepackage{amsmath,amsfonts,amssymb}
\usepackage[nosort]{cite}
\usepackage{hyperref}
\usepackage{graphicx}
\usepackage{color}
\usepackage{epsfig}
\usepackage{psfrag}	
\usepackage{tikz}
\usepackage{slashed}
\usepackage{ulem}
\usepackage[font=small,labelfont=bf,width=1.1\textwidth]{caption}
\usepackage{comment}
\usepackage[all]{xy}
\usepackage{subcaption}			% For captioning multi-panel figures

\newcommand{\sslash}{\mathbin{/\mkern-6mu/}}

\usepackage{color}

\newcommand \be{\begin{eqnarray}}
\newcommand \ee{\end{eqnarray}}

\numberwithin{equation}{section}

\DeclareMathOperator{\llangle}{\big\langle\hspace{-1.2mm}\big\langle\hspace{-.5mm}}
\DeclareMathOperator{\rrangle}{\hspace{-.5mm}\big\rangle\hspace{-1.2mm}\big\rangle}
\DeclareMathOperator{\Tr}{Tr}

\DeclareMathOperator{\sTr}{sTr}

\def\bC {\mathbb{C}}

\def\bR {\mathbb{R}}

\def\bZ {\mathbb{Z}}

\newcommand{\bea}{\begin{eqnarray}}
\newcommand{\eea}{\end{eqnarray}}
\newcommand{\beq}{\begin{equation}}
\newcommand{\eeq}{\end{equation}}
\newcommand{\bal}{\begin{equation}\begin{aligned}}
\newcommand{\eal}{\end{aligned} \end{equation}}

\newcommand{\vev}[1]{{\left< {#1} \right>}}

\newcommand{\address}[1]{\vbox{\center\em#1}}
\renewcommand{\title}[1]{\vbox{\center\huge{#1}}\vspace{5mm}}

\newcommand{\cG}{{\mathcal G}}

\newcommand{\cL}{{\mathcal L}}

\newcommand{\cN}{{\mathcal N}}
\newcommand{\cP}{{\mathcal P}}
\newcommand{\cQ}{{\mathcal Q}}

\makeatletter
\DeclareFontFamily{OMX}{MnSymbolE}{}
\DeclareSymbolFont{MnLargeSymbols}{OMX}{MnSymbolE}{m}{n}
\SetSymbolFont{MnLargeSymbols}{bold}{OMX}{MnSymbolE}{b}{n}
\DeclareFontShape{OMX}{MnSymbolE}{m}{n}{
<-6>  MnSymbolE5
<6-7>  MnSymbolE6
<7-8>  MnSymbolE7
<8-9>  MnSymbolE8
<9-10> MnSymbolE9
<10-12> MnSymbolE10
<12->   MnSymbolE12
}{}
\DeclareFontShape{OMX}{MnSymbolE}{b}{n}{
<-6>  MnSymbolE-Bold5
<6-7>  MnSymbolE-Bold6
<7-8>  MnSymbolE-Bold7
<8-9>  MnSymbolE-Bold8
<9-10> MnSymbolE-Bold9
<10-12> MnSymbolE-Bold10
<12->   MnSymbolE-Bold12
}{}
\let\llangle\@undefined
\let\rrangle\@undefined
\DeclareMathDelimiter{\llangle}{\mathopen}%
               {MnLargeSymbols}{'164}{MnLargeSymbols}{'164}
\DeclareMathDelimiter{\rrangle}{\mathclose}%
               {MnLargeSymbols}{'171}{MnLargeSymbols}{'171}
\makeatother

\begin{document}

\begin{titlepage}
\begin{center}

\vspace*{20mm}

\title{BPS Wilson loops and quiver varieties}

\vspace{10mm}

\renewcommand{\thefootnote}{$\alph{footnote}$}

Nadav Drukker%,$^1$
\footnote{\href{mailto:nadav.drukker@gmail.com}
{\tt nadav.drukker@gmail.com}}
\vskip 3mm
\address{
%$^2$
Department of Mathematics, King's College London,
\\
The Strand, London WC2R 2LS, United-Kingdom}

\end{center}

\vspace{8mm}
\abstract{
\normalsize{
\noindent
Three dimensional supersymmetric field theories have large moduli spaces of circular Wilson loops 
preserving a fixed set of supercharges. We simplify previous constructions of such Wilson loops 
and amend and clarify their classification. For a generic quiver gauge theory we identify the moduli space as a 
quotient of $\bC^m$ for some $m$ by an appropriate symmetry group. 
These spaces are quiver varieties associated to a cover of the original quiver or a 
subquiver thereof. This moduli space is generically singular and at the 
singularities there are large degeneracies of operators which seem different, but whose 
expectation values and correlation functions with all other gauge invariant operators are 
identical. 
The formulation presented here, where the Wilson loops are on $S^3$ or squashed $S^3_b$ 
also allows to directly 
implement a localization procedure on these observables, which previously required an indirect 
cohomological equivalence argument.
}}
\vfill

\end{titlepage}
%%%%%%%%%%%%%

\section{Introduction and conclusion}
\label{sec:intro}

Three dimensional conformal field theories have an intricate spectrum of line operators. 
The simplest Wilson loops mirror the $1/2$ BPS Wilson loop of $\cN\geq2$ supersymmetric Yang-Mills 
theory in 4d \cite{Gaiotto:2007qi, Drukker:2008zx, Chen:2008bp, Rey:2008bh}. 
Another construction was required to express the $1/2$ BPS Wilson loop 
in ABJM theory \cite{Drukker:2009hy, Lee:2010hk}. 
A few years ago it was realized that the possibilities of constructing 
Wilson loops are much larger, starting with $\cN=4$ quiver gauge theories, where there 
is a finite degeneracy of $1/2$ BPS operators \cite{Ouyang:2015qma, Cooke:2015ila} 
to all theories with $2\leq\cN\leq6$ with 
large moduli spaces of Wilson loops preserving four real supercharges 
\cite{Ouyang:2015iza,Ouyang:2015bmy,Mauri:2017whf,Mauri:2018fsf}. 
Beyond the Wilson loops there are further operators dubbed vortex loops 
\cite{Drukker:2008jm, Kapustin:2012iw, Drukker:2012sr}.

In Chapter~2 of a recent collaborative paper \cite{Drukker:2019bev}, Nagaoka, Probst, Tenser and 
Tr\'epanier presented a new formalism for constructing the family of $1/6$ BPS Wilson loops in ABJM theory 
and identified the moduli space as two copies of the conifold. 
Here we adapt that formalism to arbitrary theories with 
$\cN\geq2$ supersymmetry and implement it for the circular Wilson loop on the 3-sphere, 
possibly squashed $S^3_b$. In the process several new classes of operators which have not been 
identified previously are presented.

In addition to uncovering these new BPS operators, this constructive approach elucidates 
rather opaque details of previous constructions in terms of the mathematics of quivers. To 
summarize the results, the Wilson loops classification goes in two steps:
\begin{enumerate}
\item
One chooses a quiver diagram, which is 
related to the quiver of the gauge theory, but not necessarily identical to it. The 
allowed choice corresponds to and generalizes some discrete possibilities that 
arise in the solutions to the equations in the previous constructions.
\item
Given the quiver, one chooses a representation thereof, assigning numbers to the 
nodes and linear maps to the arrows. The numbers correspond to the multiplicity of 
the gauge field in the Wilson loop and the linear maps encapsulate couplings of the Wilson loop 
to the matter fields. A residual gauge symmetry (which was missed in most previous classifications) 
introduces a quotient on the linear space of maps, giving spaces known as quiver varieties.
\end{enumerate}
Rather than giving a detailed comparison to the previous works, the Wilson loops 
are constructed in the following from the ground up. The very basics of the mathematics 
of quivers, their representations and varieties are presented to make the paper 
self contained.

The construction of the Wilson loops entails certain degeneracies. Some of them are residual 
gauge transformations leading to the quotients. Beyond that, Wilson loops at fixed points of this action 
and many nearby singular orbits are actually identical as quantum operators. 
The connections and thus the holonomies are, say, upper triangular and since Wilson 
loops are traced, they do not depend on anything above the diagonal. Identifying these 
constructions leads to conical moduli spaces, like the aforementioned singular conifold.

In the next section we apply the techniques of Chapter~2 of \cite{Drukker:2019bev} to 
arbitrary theories with $\cN\geq2$ in three dimensions on $S^3$. 
We start with several examples, include the theory with one vector multiplet and 
several fundamental and/or anti-fundamental fields. From there we go 
to theories with multiple vector multiplets.

The case 
of the squashed sphere and theories with fields of non-canonical dimensions are 
studied in Appendix~\ref{app}. Some of this analysis has already been done in \cite{Mauri:2018fsf}, 
but we use a different formalism and generalize their constructions. We note that 
using the language of off-shell $\cN=2$ supersymmetry allows to perform supersymmetric 
localization immediately without resorting to a chomological-equivalence argument 
\cite{Kapustin:2009kz, Hama:2010av,Hama:2011ea}.

An important ingredient in the $1/2$ BPS Wilson loop of ABJM theory and in most of 
the other operators is the coupling to Fermi fields. In the original papers this coupling 
had a rather subtle path dependence, which was reinterpreted in \cite{Drukker:2019bev} as 
a constant shift in the bosonic connection, simplifying the expressions. The analysis here 
elucidates the origin of this shift as arising from the symmetry algebra and related to 
the curvature of $S^3$ (and the background vector field on the squashed sphere). 
This is the last term in \eqref{Q2}, and it was already noticed in Chapter~2 of \cite{Drukker:2019bev}, 
in the context of the circular Wilson loop in $\bR^3$. These shifts are also crucial in 
the construction of the quiver representing the Wilson loops and different shifts are 
encoded in different (graded) quiver diagrams.

The moduli spaces are studied in Section~\ref{sec:moduli}, again starting with several simple examples. 
The role of the shifts in modifying the original quiver are presented and the subsequent map between 
Wilson loop data and that of quiver representations is then explained.

Most of the discussion in this paper is classical, except where we point out how localization 
can be applied. It is an interesting question to 
verify to what extent the statements made here are subject to quantum corrections, 
as with very little supersymmety one would expect them to arise.

A natural avenue to address that is by viewing Wilson loops as defect CFTs. 
This leaves many questions on the anomalous dimensions of insertions into the Wilson 
loop and especially the relation between the moduli spaces 
found here and the Zamolodchikov metric of the defect CFTs.

\section{BPS Wilson loops and quiver representations}
\label{sec:WL}
\subsection{Wilson loop from vector multiplet}

Any 3d theory with $\cN\geq2$ supersymmetry and a vector multiplet 
$(A_\mu, \lambda, \bar\lambda, \sigma, D)$ has a BPS Wilson loop 
of the form \cite{Gaiotto:2007qi}
\beq
\label{bos}
W=\Tr\cP \exp\oint \left(i A_\mu\dot x^\mu +\sigma|\dot x|\right)d\tau\,.
\eeq
We consider the Euclidean theory on $S^3$ of radius $R$, where the path is a great circle in the 
direction of the dreibein $e^1=R\,d\varphi$. One can just as well take 
a circle in flat $\bR^3$, or as done in the appendix, the squashed sphere. 

Using the variations in \eqref{delta-vec}, it is easy to show that 
the Wilson loop is invariant under supersymmetry as long as the 
independent parameters $\epsilon$ and $\bar\epsilon$ satisfy
\beq
\label{projector}
(\gamma_1-1)\epsilon=(\gamma_1+1)\bar\epsilon=0\,.
\eeq
On $S^3$ this restricts the chirality of the supercharges to 
$\epsilon^1$ and $\bar\epsilon^2$, while on $S^3_b$, it 
restricts the loop to a particular circle at $\vartheta=0$. Denoting the 
corresponding supercharges $Q$ and $\bar Q$, these two annihilate the Wilson loop.
In the following we use the two linear combinations of them
$\cQ_\pm=Q\pm\bar Q$.

\subsection{Wilson loop with matter}
Let us assume that in addition to the vector multiplet the theory has 
$n$ fundamental chiral fields $(\phi^i, \psi^i, F^i)$ with $i=1,\cdots, n$ 
and their conjugates. Here we take that the fields have canonical dimensions 
$(1/2, 1, 3/2)$ respectively, which is guaranteed for $\cN>2$. The 
case of $\cN=2$ with non canonical dimensions is presented in the appendix.

Using the SUSY transformation of 
\cite{Kapustin:2009kz, Hama:2010av,Hama:2011ea, Marino:2011nm, Drukker:2012sr}, 
summarized in the appendix, we have that the scalar fields satisfy 
\eqref{Q2}
\beq
\label{deltasquare}
R\cQ_+^2\phi=i\partial_\varphi\phi-A_\varphi\phi+iR\sigma\phi-\frac{1}{2}\phi\,,
\qquad
R\cQ_+^2\bar\phi=i\partial_\varphi\bar\phi+\bar\phi A_\varphi-iR\bar\phi\sigma+\frac{1}{2}\bar\phi\,.
\eeq
To account for the factor of $1/2$, it is natural to shift the connection in \eqref{bos} and 
for the purpose of coupling to the chiral fields we package the bosonic loop in a $2\times2$ 
block diagonal structure as
\beq
\label{L0}
W+1=-\sTr\cP \exp\oint i \cL_0\,|\dot x| \,d\tau\,,
\qquad
\cL_0=\begin{pmatrix}
A_\mu\frac{\dot x^\mu}{|\dot x|}-i\sigma+\frac{1}{2R}\ &
0\\
0&
0
\end{pmatrix}.
\eeq
The extra 1 on the left hand side of \eqref{L0} accounts for the contribution from the trivial 
$1\times1$ block. From now on we absorb this and the overall sign in front of the supertrace 
into $W$. 
We could assign the constant piece in the connection to the lower-right block, as $-1/2R$, 
or following \cite{Drukker:2019bev} as $\pm1/4R$ each in the upper and lower blocks. 
The changes amounts to adding a constant to 
$\cL_0$, or multiplying the Wilson loop by the phase $-1$ or $-i$ respectively.

We now place 
the chiral and anti-chiral fields into the off-diagonal entries of $(N|1)$ odd-supermatrices, 
where the off-diagonal entries are Gra\ss mann even.
\beq
\label{G}
\cG_{u,\bar u}=G_u+\bar G_{\bar u}\,,
\qquad
G_u=\begin{pmatrix}0& u_i\phi^i\\0&0\end{pmatrix},
\qquad
\bar G_{\bar u}=\begin{pmatrix}0&0\\\bar u^i\bar\phi_i&0\end{pmatrix},.
\eeq
$u^i$ and $\bar u_i$ are arbitrary complex vectors (not necessarily complex conjugates). 
With this we can compactly write \eqref{deltasquare} as
\beq
\label{delta2G}
\cQ_+^2\cG_{u,\bar u}
=i{\mathfrak D}_0\cG_{u,\bar u}
\equiv\frac{i}{R}\partial_\varphi\cG_{u,\bar u}-\left[\cL_0,\cG_{u,\bar u}\right].
\eeq
We use $\cG_{u,\bar u}$ to deform the Wilson loop to
\beq
\label{Lubaru}
W_{u,\bar u}=\sTr\cP \exp\oint i \cL_{u,\bar u}\,|\dot x| \,d\tau\,,
\qquad
\cL_{u,\bar u}=\cL_0-i\cQ_+\cG_{u,\bar u}
+\cG_{u,\bar u}^2\,.
\eeq

\begin{figure}[t]
\begin{centering}
\psfrag{N}{\,\large$N$}
\psfrag{n}{\large$n$}
\begin{minipage}[b]{7cm} 
\center\raisebox{2pt}{\includegraphics[width=5cm]{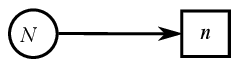}}
\subcaption{Quiver diagram representing a gauge theory with one $SU(N)$ vector 
multiplet and $n$ fundamental chirals.\\
\label{Qfund}}
\end{minipage}
\qquad
\begin{minipage}[b]{7cm} 
\psfrag{N}{\,\large$1$}
\center\includegraphics[width=5cm]{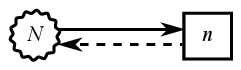}
\subcaption{Quiver diagram representing a Wilson loop with one copy of the 
gauge field shifted by $1/2R$ and arrows for the couplings $u_i$ and $\bar u^i$.
\label{QfundW}}
\end{minipage}
\caption{\label{fig:fundW}We represent the Wilson loop \eqref{Lubaru} in the theory with 
the quiver (a) by the quiver diagram (b). The squiggly circle indicates the shift by $1/2R$.}
\end{centering}
\end{figure}

Under supersymmetry transformations
\beq
\cQ_+\cL_{u,\bar u}
=\cQ_+\cL_0-i\cQ_+^2\cG_{u,\bar u}
+\left\{\cQ_+\cG_{u, \bar u},\cG_{u,\bar u}\right\}
%\\&
=\mathfrak{D}_0\cG_{u,\bar u}+\left\{\cQ_+\cG_{u,\bar u},\cG_{u,\bar u}\right\}.
\eeq
$\mathfrak{D}_0$ is the covariant derivative along the loop with the connection $\cL_0$.
We can replace it by $\mathfrak{D}_{u,\bar u}$, with the new connection
\beq
\label{total-der}
\cQ_+\cL_{u,\bar u}
=\mathfrak{D}_{u,\bar u}\cG_{u,\bar u}
\equiv
\frac{1}{R}\partial_{\varphi}\cG_{u,\bar u}
+i\left[\cL_0,\cG_{u,\bar u}\right]
+\left\{\cQ_+\cG_{u, \bar u},\cG_{u,\bar u}\right\}
+i\left[\cG_{u,\bar u}^2,\cG_{u,\bar u}\right].
\eeq
This is a total derivative (or a field-valued supergauge transformation), so the Wilson loop $W_{u,\bar u}$ 
is invariant under this supersymmetry transformation. Note that guaranteeing cancelation of this term upon 
integration is the reason for the supertrace  in the definition of the Wilson loop \eqref{Lubaru}. Also, the 
fact that one of the commutators is replaced with an anti-commutator is due to $\cQ_+\cG_{u,\bar u}$ being 
Gra{\ss}mann odd. The cancelation of the total derivative terms and all the signs were checked carefully in 
\cite{Drukker:2009hy} in the case of ABJM theory and the argument carries over.

A quiver diagram representing this Wilson loop is illustrated in Figure~\ref{QfundW}, 
where the shift of the gauge connection 
in \eqref{L0} is represented by the squiggly circle. The notation and subsequent classification 
of loop operators are explained in Section~\ref{sec:gen}.

To check invariance under the second supercharge, we note that
\beq
\cQ_-G_{u}=-\cQ_+G_{u}\,,
\quad
\cQ_-\bar G_{\bar u}=\cQ_+\bar G_{\bar u}\,,
\quad
\cQ_-^2G_{u}=-\cQ_+^2G_{u}\,,
\quad
\cQ_-^2\bar G_{\bar u}=-\cQ_+^2\bar G_{\bar u}\,,
\eeq
which gives
\beq
\label{1/4}
\cQ_-\cL_{u,\bar u}
=\mathfrak{D}_{u,\bar u}(G_{u}-\bar G_{\bar u})
-2\cQ_+(G_u^2-\bar G_{\bar u}^2)
-2i\left[G_u^2+\bar G_{\bar u}^2,G_u-\bar G_{\bar u}\right].
\eeq
For the last two terms to vanish we need to require that both 
$G_u$ and $\bar G_{\bar u}$ are nilpotent of index 2, which 
indeed is the case in \eqref{G}. In other theories this may be a non-trivial constraint. 
If $G$ or $\bar G$ do not square to zero, we end 
up with $1/4$ BPS loops. With some changes of signs in \eqref{Lubaru} we can 
construct $1/4$ BPS loops invariant under $\cQ_-$ instead of $\cQ_+$.

The next ingredient we want to consider are anti-fundamental chiral mulitplets $\tilde\phi^i$. 
The analog of \eqref{deltasquare} is
\beq
\label{tilde}
R\cQ_+^2\tilde\phi=i\partial_\varphi\tilde\phi+\tilde\phi A_\varphi-iR\tilde\phi\sigma-\frac{1}{2}\tilde\phi\,,
\qquad
R\cQ_+^2\bar{\tilde\phi}
=i\partial_\varphi\bar{\tilde\phi}-A_\varphi\bar{\tilde\phi}+iR\sigma\bar{\tilde\phi}+\frac{1}{2}\bar{\tilde\phi}\,.
\eeq
$\tilde\phi$ transforms similarly to $\bar\phi$, expect for the change in sign of the last term. It is 
then possible to construct matter Wilson loops with $\tilde\phi$ and $\bar{\tilde\phi}$, by replacing 
$1/2R$ in $\cL_0$ with $-1/2R$.

\begin{figure}[t]
\begin{centering}
\psfrag{N}{\,\large$N$}
\psfrag{n}{\large$n$}
\psfrag{nt}{\large$\tilde n$}
\begin{minipage}[b]{7cm} 
\center\raisebox{2pt}{\includegraphics[width=7cm]{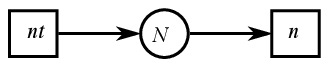}}
\subcaption{Quiver diagram of a gauge theory with $SU(N)$ vector 
multiplet, $n$ fundamental chirals and $\tilde n$ antifundamental chirals.
\label{Qantifund}}
\end{minipage}
\qquad
\begin{minipage}[b]{7cm} 
\psfrag{N}{\ \large$1$}
\center\includegraphics[width=7cm]{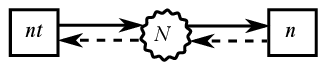}
\subcaption{Quiver diagram for the Wilson loop with 
one gauge field shifted by $1/2R$ and couplings to all the chirals and antichirals.\label{QantifundW}}
\end{minipage}
\caption{\label{fig:antifundW}The Wilson loop \eqref{1/2,0,-1/2} in the theory with 
the quiver (a) is represented by the quiver diagram (b). It is 
$1/4$ BPS because there is a solid arrow going into the squiggly node. 
For the shift to decrease by $1/2R$ along a solid arrow, $\tilde n$ node 
should have shift $1/R$ as in \eqref{1/2,0,-1/2}. Alternatively, if we 
gague transform to \eqref{lat}, the solid line pointing into 
a squiggly circle indicates that the fields have explicit phases.}
\end{centering}
\end{figure}

If we want to include both fundamental $\phi$ and anti-fundamental $\tilde\phi$ fields, we can 
use a $3\times3$ structure which forms an $(N|2)$ supermatrix
\beq
\label{1/2,0,-1/2}
\cL_0=\begin{pmatrix}\frac{1}{R}&0&0\\
0&A_\mu\frac{\dot x^\mu}{|\dot x|}-i\sigma+\frac{1}{2R}&0\\
0&0&0\end{pmatrix},
\quad
G_{v;u}=\begin{pmatrix}0& v_j\tilde \phi^j&0\\0&0&u_i\phi^i\\0&0&0\end{pmatrix},
\quad
\bar G_{\bar v;\bar u}=\begin{pmatrix}0&0&0\\\bar v^j\bar{\tilde\phi}_j&0&0\\0&\bar u^i\bar\phi_i&0\end{pmatrix}.
\eeq
Constructing $\cL_{v,\bar v;u,\bar u}$ out of these ingredients as before leads to an operator invariant 
under $\cQ_+$, but since $G_{v;u}$ and $\bar G_{\bar v;\bar u}$ do not square to zero, it is not invariant under 
$\cQ_-$, so only $1/4$ BPS.

Another approach is to incorporate an explicit phase $\tilde\phi\to e^{-i\varphi}\tilde\phi$, which now 
transforms under $\cQ_+^2$ exactly like $\bar\phi$. We can then construct the $2\times2$ superconnection 
out of $\cL_0$ as in \eqref{L0} and
\beq
\label{lat}
G_{v;u}=\begin{pmatrix}
0&\ u_i\phi^i\\e^{i\varphi}v_j\tilde\phi^j&0
\end{pmatrix},
\qquad
\bar G_{\bar v;\bar u}=\begin{pmatrix}
0&e^{-i\varphi}\bar v^j\bar{\tilde\phi}_j\\\bar u^i\bar\phi_i\ &0\\
\end{pmatrix}.
\eeq
Again we find that $G_{v;u}$ and $\bar G_{\bar v;\bar u}$ are not 
nilponent, so the resulting Wilson loop is $1/4$ BPS, not $1/2$. These Wilson loops are analogous 
to the ``latitude'' Wilson loops of ABJM \cite{Cardinali:2012ru,Bianchi:2018bke}, related to the 
$1/4$ BPS circular Wilson loops of $\cN=4$ supersymmetric Yang-Mills theory in 4d \cite{Drukker:2006ga}. 

The quiver diagram for the theory with both fundamental and antifundamental chirals is in Figure~\ref{Qantifund}. 
The Wilson loops that couple to all the fields is represented by the quiver in Figure~\ref{QantifundW}. 
The solid arrows represent $G_u$ and the fact that $G_u^2\neq0$ is evident from the pair 
of consecutive arrows, so this is the graphical condition distinguishing Wilson loops 
preserving only $\cQ_+$ or also $\cQ_-$.

\begin{figure}[t]
\begin{centering}
\hskip-2cm
\psfrag{w}{\Large\color[rgb]{.211,.047,.375}$e^{i\int (A-(i\sigma-\phi\bar\phi-\frac{1}{2R})|\dot x|d\tau)}$}
\psfrag{p}{\color[rgb]{0.371,0.145,0}$\psi$}
\psfrag{pb}{\color[rgb]{0.371,0.145,0}$\bar\psi$}
\psfrag{f}{\Large\color[rgb]{0.0195,0.398,0.2266}$e^{i\int\bar\phi\phi|\dot x|d\tau}$}
\includegraphics[width=9cm]{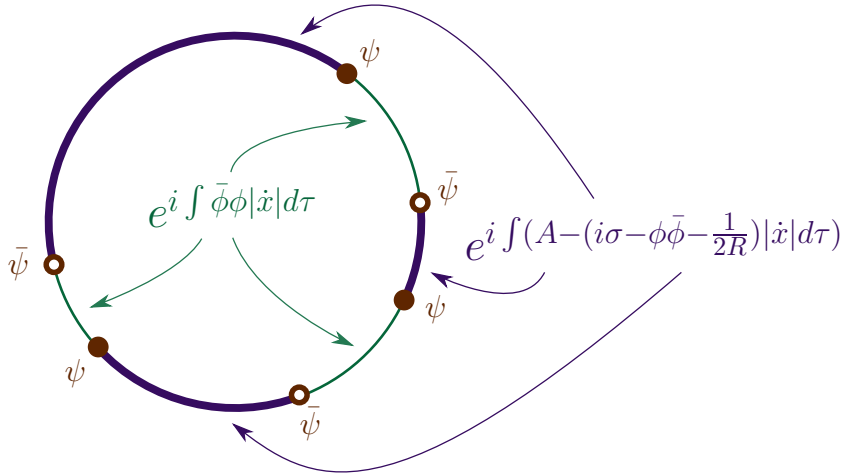}
\caption{\label{fig:open}A contribution to the Wilson loop made of the superconnection in 
\eqref{Lopen} when expanded to sixth order in the fermions, represented here as small 
circles. At these locations the operator alternates between 
open Wilson lines with proper connections (thick arcs) and integrals of the scalar bilinear 
(thin arcs).}
\end{centering}
\end{figure}

\subsection{Explicit expressions}
To write the Wilson loops explicitly, we use the action of $Q$ and $\bar Q$ in \eqref{delta-chi}. 
Equation \eqref{Lubaru} gives
\beq
\label{Lopen}
\cL_{u,\bar u}
=\cL_0
+\begin{pmatrix}
u_i\bar u^j\phi^i\bar\phi_j&
-iu_i\psi_2^i \\
-i\bar u^j\bar\psi_{j1}&
u_i\bar u^j\bar\phi_j\phi^i
\end{pmatrix}.
\eeq
Here $\psi_2$ and $\bar\psi_1$ are spinor eigenstates of $\sigma_1$, see \eqref{projector}.
The Wilson loop is the supertrace of the holonomy of this superconnection.

It is worthwhile to pause and mull over this formal expression. The top left corner of 
the supermatrix is valued in the adjoint of $SU(N)$, as in a usual Wilson loop. The other 
even entry, at the bottom right, essentially $\bar\phi\phi$, is an $SU(N)$ singlet. This part is not 
required for gauge invariance, but is crucial to guarantee supersymmetry. The odd entries, 
with $\psi^i$ and $\bar\psi_j$ transform in the fundamental and anti-fundamental of $SU(N)$. 
Such fields serve as endpoints of open Wilson lines. So in terms of $SU(N)$ Wilson loops, 
the $(N|1)$ Wilson loop is a linear combination of many Wilson lines. One is a 
closed loop with the modified connection in the upper left corner. The others are collections 
of open arcs with the fermions as start and endpoints and this modified connection between them. 
The gaps between the open arcs are filled by the trivial Wilson loops made of the singlet 
component. See the illustration in Figure~\ref{fig:open}.

\begin{figure}[t]
\begin{centering}
\psfrag{N}{\,\large$N$}
\psfrag{p}{\large$1$}
\psfrag{q}{\large$1$}
\begin{minipage}[b]{5cm} 
\center\raisebox{2pt}{\includegraphics[width=2.7cm]{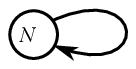}}
\subcaption{Quiver diagram of a gauge theory with $SU(N)$ vector 
multiplet and adjoint chiral.\\ }
\end{minipage}
\quad
\begin{minipage}[b]{5cm} 
\center\includegraphics[width=5cm]{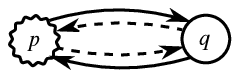}
\subcaption{A cover of (a) with one unshifted and 
one shifted gauge field and all possible matter couplings.}
\end{minipage}
\quad
\begin{minipage}[b]{5cm} 
\center\raisebox{4pt}{\includegraphics[width=5cm]{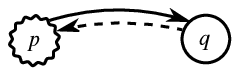}}
\subcaption{\label{fig:adj-1/2}A subquiver of (b) with matter couplings removed so the 
solid arrows only point out of the squiggly circle.}
\end{minipage}
\caption{\label{fig:adjoint}Constructing Wilson loops with adjoint chiral matter in the 
theory (a) requires doubling the quiver, with one shifted and one unshifted node (b,c). 
With all the matter couplings, the resulting Wilson loop (b) is $1/4$ BPS. 
The subquiver (c) represents \eqref{adjoint}, with only chirals in the upper right block 
and only anti-chirals in the bottom left, so is 1/2 BPS. }
\end{centering}
\end{figure}

Before attempting to describe the most general Wilson loop arising in this way, as is done 
in the next section, here is another example for a field theory with one vector multiplet 
and several adjoint chirals. We view the matter field as if it is in the bi-fundamental of two copies 
of the gauge group, so define the doubled $2\times2$ structure (or $(N|N)$ superconnection)
\beq
\label{adjoint}
\cL_{u,\bar u}
=\begin{pmatrix}
A_\mu\frac{\dot x^\mu}{|\dot x|}-i\sigma+u_i\bar u^j\phi^i\bar\phi_j+\frac{1}{2R}&
-iu_i\psi^i_2 \\
-i\bar u^j\bar\psi_{j1}&
A_\mu\frac{\dot x^\mu}{|\dot x|}-i\sigma-u_i\bar u^j\bar\phi_j\phi^i
\end{pmatrix}.
\eeq
Note that the two diagonal blocks have the same gauge field and $\sigma$, but a different 
combination of the scalars and a different constant shift (which cannot be gauged away by 
a single valued gauge transformation). The Wilson loop made out of this 
connection is in a representation of $SU(N|N)$ and its quiver is in Figure~\ref{fig:adj-1/2}.

\subsection{General theories and quiver representations}
\label{sec:gen}

\begin{figure}[t]
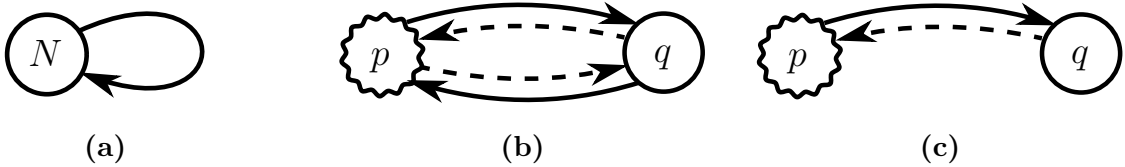

\begin{centering}
\psfrag{N}{\,\large$N$}
\psfrag{p}{\large$p$}
\psfrag{q}{\large$q$}
\begin{minipage}[b]{5cm} 
\center\raisebox{2pt}{\includegraphics[width=2.7cm]{adjoint1.eps}}
\subcaption{}
\end{minipage}
\quad
\begin{minipage}[b]{5cm} 
\center\includegraphics[width=5cm]{adjoint2.eps}
\subcaption{}
\end{minipage}
\quad
\begin{minipage}[b]{5cm} 
\center\raisebox{4pt}{\includegraphics[width=5cm]{adjoint3.eps}}
\subcaption{}
\end{minipage}
\caption{\label{fig:adjointpq}Quiver diagrams for more general Wilson loops 
in the theory with adjoint matter (a). With $p$ copies of the shifted gauge field 
and $q$ copies unshifted, on the diagonal of $\cL_0$, allowing for all matter 
couplings (b) gives $1/4$ BPS Wilson loops, or restricting to only half 
the coupling (c) gives $1/2$ BPS loops.}
\end{centering}
\end{figure}

Following the examples above, we can construct a very general Wilson loop as follows.

The first step is the construction of the block-diagonal superconnection $\cL_0$ with 
any number of copies of shifted or unshifted gauge fields (with the appropriate $\sigma$). 
We use gauge transformations as in \eqref{lat} to make all the shifts $0$ or $1/2R$. 
The vector multiplets are nodes of the original quiver and any vector field appearing 
unshifted in $\cL_0$ is represented again as a circle, with its degeneracy written inside. 
A shifted vector field is indicated by a wiggly circle, 
again decorated by its degeneracy. Not all nodes of the original quiver have to be 
represented in the new quiver, but some may be doubled, as in the 
example in Figure~\ref{fig:adjoint}. The Wilson loop in the same theory 
with more copies of the gauge fields ($p$ shifted and $q$ unshifted) is 
represented by the quivers in Figure~\ref{fig:adjointpq}.

Given $\cL_0$ we construct the off-diagonal matrices $G$ and $\bar G$ with the 
matter fields with entries only connecting shifted and unshifted entries in 
$\cL_0$. The shifts endow all the matrices with $\bZ_2$ gradings. We can split $\cL_0$ into 
a shifted and unshifted blocks in which case $G$ and $\bar G$ are in the two 
complementary blocks. The resulting superconnection is then clearly a supermatrix.

In the quiver representing the Wilson loop, the non-zero entries in $G$ are represented by solid 
arrows, and of course mirror the chiral multiplets in the original quiver. 
$\bar G$, which includes the anti-chiral fields, is represented by dashed arrows. These 
matrices incorporate couplings similar to $u_i$ above, furnishing the Wilson loops 
with continuous parameters. They are rectangular matrices, \emph{i.e.}, linear maps 
between spaces associated with the nodes, forming a quiver represntation.

We thus find a construction involving a discrete choice of quiver, which is a sub-quiver of 
a double cover of the original quiver, and continuous parameters. In Section~\ref{sec:moduli}, 
we study the continuous parameters in greater detail and distinguish the moduli 
spaces of $1/2$ BPS and $1/4$ BPS Wilson loops.

\begin{figure}[t]
\begin{centering}
\psfrag{N}{\,\large$N$}
\psfrag{M}{\large$M$}
\begin{minipage}[b]{5cm} 
\center\includegraphics[width=5cm]{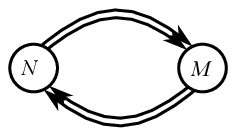}
\subcaption{}
\end{minipage}
\quad
\psfrag{N}{\large$\ p$}
\psfrag{M}{\large$\ q$}
\begin{minipage}[b]{5cm} 
\center\raisebox{20pt}{\includegraphics[width=5cm]{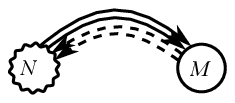}}
\subcaption{}
\end{minipage}
\quad
\begin{minipage}[b]{5cm} 
\center\includegraphics[width=5cm]{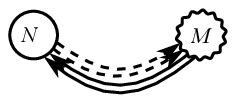}
\subcaption{}
\end{minipage}
\caption{\label{fig:ABJ} (a) The quiver diagram of ABJM theory. 
Shifting the left (b) or right (c) nodes matches the 
class I and class II Wilson loops in the classification of 
\cite{Ouyang:2015iza,Ouyang:2015bmy,Mauri:2017whf,Mauri:2018fsf}.}
\end{centering}
\end{figure}

As a further example, consider ABJ(M) theory, as illustrated in Figure~\ref{fig:ABJ}. The 
gauge theory has two nodes, and to construct Wilson loops we need to grade it. There 
are two possibilities, indicated in the right two diagrams there 
(class I and II in the classification of \cite{Ouyang:2015iza,Ouyang:2015bmy,Mauri:2017whf,Mauri:2018fsf}). 
The middle diagram represents 
Wilson loops where the gauge field of the left node is shifted by $1/2R$ and the second node 
is unshifted. The first appears $q$ times in the diagonal of $\cL_0$ and the second 
$p$ times. The right most diagram has the shift on the other gauge fields. Note that we 
can view the union of the two diagrams on the right as a double cover of the original quiver, 
where each node has a shifted and unshifted copy and we retain only the solid arrows 
out of the shifted (squiggly) nodes. Since the cover is disconnected, we get two 
options, unlike the case when the quiver has odd cycles as in Figure~\ref{fig:adjointpq}.

In the graphical representation of the Wilson loops as quivers, the arrows represent 
the scalar couplings. As mentioned around \eqref{1/4}, while $\cQ_+$ is preserved 
with coupling to any of the matter fields, for the loop 
to be invariant under $\cQ_-$ we have to impose the nilpotency condition 
$G^2=\bar G^2$. $G$ are the solid arrows and $G^2$ are two consecutive solid 
arrows. Thus the graphical condition for $1/2$ BPS loops is that all nodes have either all 
outgoing solid arrows (the squiggly circles) or all ingoing (unsquiggly), and likewise for dashed arrows.

More precisely, the dashed arrows need to point in the opposite direction, otherwise 
also $(G+\bar G)^2=0$, rendering the last term in \eqref{Lubaru} trivial. Formally this 
operator is $1/2$ BPS 
(it was called class III and class IV in \cite{Ouyang:2015iza,Ouyang:2015bmy,Mauri:2017whf,Mauri:2018fsf}), 
but $\cL$ in such cases is (block) upper/lower triangular with the same diagonal as 
$\cL_0$. As explained in the next section, the Wilson loops constructed from such a connection 
are identical as quantum operators to those made from $\cL_0$.

\begin{figure}[t]
\begin{centering}
\psfrag{N}{\large$\ p$}
\psfrag{M}{\large$\ q$}
\center\includegraphics[width=5cm]{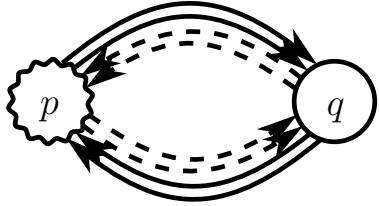}
\caption{\label{fig:ABJMW}Quiver representation of the $1/4$ BPS Wilson loop 
in ABJ(M) theory.}
\end{centering}
\end{figure}

The two quiver diagrams on the right of Figure~\ref{fig:ABJ} are the possible Wilson loops 
preserving $\cQ_+$ and $\cQ_-$ (which makes them $1/6$ BPS with respect to the 
$\cN=6$ of ABJ(M)). If we include the other couplings, giving a total of $8pq$ parameters, 
we have $1/12$ Wilson loops represented by the quiver diagram in Figure~\ref{fig:ABJMW}. 
Note that the solid arrows going into the squiggly circle violate the grading rule in 
\eqref{1/2,0,-1/2}, so we are actually in the setting of \eqref{lat}, where we included 
explicit angle dependent phases and these Wilson loops 
are generalizations of the ``latitude'' Wilson loops of \cite{Cardinali:2012ru,Bianchi:2018bke}.

\subsection{More cases}
\label{sec:more}
Let us discuss three extra ingredients: squashing the sphere, theories with non-canonical 
dimensions and the theory on $\bR^3$.

The construction for the squashed sphere $S^3_b$, presented in the appendix, follows the 
exact same structure and the only modification is changing $1/2R$ to $(1+b^2)/4bR$. The 
classification of $1/2$ BPS loops, based on $\bZ_2$ graded quivers (with squiggly and 
unsquggly nodes) which are a cover 
and/or subquiver of the original gauge theory quiver remains the same, as we allow 
only the two shifts 0 and $(1+b^2)/4bR$.

The classification of $1/4$ BPS loops is different, since now a double shift to 
$(1+b^2)/2bR$, cannot be gauged away, as it is not an integer multiple of $1/R$. 
This excludes the latitude Wilson loops as in \eqref{lat}, but the construction in 
\eqref{1/2,0,-1/2} is still valid. If $b+1/b$ is rational, then after several shifts we 
could make a gauge transformation back to 0, but otherwise we need to rely on a 
$\bZ$ grading of the quiver, which exists if it is a tree. If the quiver has loops, 
we need in principle to take an infinite cover of it, to get a $\bZ$ graded quiver 
and then base the Wilson loop on a finite subquiver of it. As usual there 
is also the doubling of arrows representing the anti-chiral fields.

In the case of theories with $\cN=2$ supersymmetry, the dimensions of the chiral fields 
are not protected by supersymmetry and may vary under renormalization group flow. 
To account for this, one can assign them non-canonical dimensions from the onset, 
where the scalars in the chiral multiplet have dimension 
$\Delta$ instead of $1/2$. In this case the shift should be $\Delta/R$. If all the chirals 
have the same dimension, the discussion remains as in the squashed $S^3_b$ case, 
but if they are different, then the shifts are not all multiples of a basic shift.

To get $1/2$ BPS loops we need the same types of quivers as before. with each node 
either having all ingoing solid arrows or all outgoing solid arrows. generically, if all dimensions 
are different, only one arrow is allowed per pair of nodes and if more exist, we need 
to take a cover of the quiver to account for that. A simple illustration is the theory with 
one vector and $n$ fundamental chirals. Instead of the single flavor node with $v^i$ in the 
vector space $\bC^n$, we have $n$ flavor nodes, each with a different shift and a 
single $v^i$ coupling. In the next section the moduli spaces of the loop operators are 
presented. In the case of a single shift, it is $(\bC^n)^2\sslash\bC^*_{-1,1}$, 
while with different shifts it turns out to be only $\bC^n$.

The $1/4$ BPS loops are based on the same infinite cover of the quiver without the 
requirement of nodes being all ingoing or outgoing. Again one cannot generically 
perform gauge transformations to get loops like in \eqref{lat}, but rather use constructions as in 
\eqref{1/2,0,-1/2}.

Note that in all of these cases, both with $S^3_b$ and generic $\Delta$, we can still 
usually assign a grading to the cover of the quiver based on the distance of a node from 
a fixed one. This integer grading does not necessarily match the shifts, but it still 
guarantees that $\cL$ is a supermatrix. This is violated if we can form odd loops, 
when $b+1/b$ is rational or when three dimensions align as in 
$\Delta_1+\Delta_2=\Delta_3$.

Projecting from $S^3$ to $\bR^3$, the Wilson loop can become a circle or an infinite 
straight line. For the circle, one replaces $R$ in the shifts with the radius of the circle 
as in \cite{Drukker:2019bev}, so the discussion remains identical to $S^3$.
In the case of the straight line, taking $R\to\infty$ leads all the shifts to vanish. 

In this case there is no need to grade the quiver or take its cover (just including the 
dashed arrows). There are no restrictions on constructing $1/4$ BPS loops, and the supersymmetry 
is doubled if $G^2=\bar G^2=0$, as usual. As already pointed out in \cite{Mauri:2018fsf}, a Wilson line for a 
triangular quiver need not satisfy the $\bZ_2$ grading and $\cL$ is not a supermatrix.

In theories with $\cN\geq4$ there are Wilson loops preserving more than the minimal 
set of supercharges. In the present formulation it is hard to identify the points of 
enhanced supersymmetry on the moduli space, as it relates to 
other details of the theory like the superpotential. In the case of ABJM an argument 
based on enhanced $SU(3)$ symmetry was presented in \cite{Drukker:2019bev}. 
Also the fermionic latitudes of ABJM preserve 2 complex supercharges, while 
the construction here guarantees only one.

A last point we have not touched upon yet are the representation of the Wilson loops and 
in all the expressions above the supertrace is assumed to be in the fundamental 
representation of the large matrix. One could perform the trace, of course also in 
higher dimensional representations. We have also not discussed when the different 
constructions are really different or are reducible. This question should be 
answered by whether the quiver representation is irreducible, as discussed 
in one example around \eqref{2N|1} below.

\subsection{Localization}
Since the constructions presented above relies on $\cN=2$ off-shell supersymmetry on the sphere, 
for all the $1/2$ BPS Wilson loops we can immediately apply supersymmetric localization in this formalism. 
The bosonic Wilson loop was already localized in \cite{Kapustin:2009kz} and the generalization 
to the loops coupled to the matter fields requires no further work, since at the localization locus 
the matter fields (as well as the gauge field) vanish. We are left with the field 
$\sigma$, fixed to a constant, so the operators are identical to the usual bosonic 
loops, or combination thereof in the extra even blocks of the superconnection. 
In the case of the theory with one vector and $n$ fundamental chirals, 
the expectation value of the Wilson loop is then given by the matrix model
\beq
\vev{W_{u,\bar u}}
=\vev{-\sTr e^{\left(\begin{smallmatrix}2\pi\sigma+\pi i&0\\0&0\end{smallmatrix}\right)}}_\text{M.M.}
=\vev{\Tr e^{2\pi\sigma}}_\text{M.M.}+1\,.
\eeq
M.M. represents matrix model calculations, the usual result of the localization procedure.

Note that we have made no reference to the action of the field theory, and the entire discussion 
goes through as long as it is supersymmetric. We can have Yang-Mills and chiral actions, 
that do not effect the matrix model. The Chern-Simons action, Fayet-Iliopoulos and mass 
terms do appear in the matrix model action, as usual 
\cite{Kapustin:2009kz, Hama:2010av,Hama:2011ea}.

The cases that are $1/4$ BPS require new techniques to perform localization with only $\cQ_+$. For 
the latitude of ABJM theory, a matrix model was proposed in \cite{Bianchi:2018bke}. It is 
possible to come up with generalizations of this proposal for all theories, but we do not pursue that here.

\section{Moduli spaces}
\label{sec:moduli}

In the example of a vector with $n$ fundamental chirals and an $(N|1)$ superconnection, the 
Wilson loop $W_{u,\bar u}$ \eqref{Lubaru} 
is defined in terms of two complex $n$-component vectors $u^i$ and $\bar u_i$, 
but it is invariant under the (constant) gauge transformation
\beq
\label{gauge}
\cL_{u,\bar u}
\to
\begin{pmatrix}1&0\\0&1/x\end{pmatrix}
\cL_{u,\bar u}
\begin{pmatrix}1&0\\0&x\end{pmatrix}
=
\cL_{xu,\bar u/x}\,.
\eeq
We find the equivalence relation $(u,\bar u)\sim(xu,\bar u/x)$, giving the moduli space
$(\bC^{n})^2\sslash\bC^\star_{1,-1}$, where the subscripts represents the weights of this action 
on each copy of $\bC^n$.

To get a nice space out of this quotient requires either further identifications or some resolution. 
For example consider $\bar u=0$, where $\bar G=0$, so $\cL_{u,0}$ is upper 
triangular and the diagonal pieces are the same as $\cL_0$. 
Any product of such a superconnection is still triangular so the Wilson loop operator 
that we get from supertracing such products is identical to the original Wilson loop without 
matter. Their expectation value and correlators with any other operators are identical. 

Based on this, we should identify the $\bar u=0$ subspace as well as that of 
$u=0$ with the origin (the Wilson loop with $\cL_0$). An alternative description of this space is that of 
complex $n$ dimensional matrices of rank one. For $n=1$ this is simply a copy 
of $\bC$ represented by the product $\bar u u$. For $n=2$ we have the four coordinates 
$P=u_1\bar  u^1$, $R=u_1 \bar u^2$, $S=u_2\bar  u^1$, $T=u_2\bar  u^2$ satisfying 
$PT-RS=0$ inside $\bC^4$, which is the equation for the singular conifold.

As already pointed out in Chapter~2 of \cite{Drukker:2019bev}, 
the moduli space in the case of the $1/6$ BPS Wilson loops in ABJM theory is two 
copies of the same singular confold. Indeed the discussion above carries over to the 
case of the two quiver diagrams on the right of Figure~\ref{fig:ABJ} for $p=q=1$, 
each giving one copy of the conifold.

\begin{figure}[t]
\begin{centering}
\psfrag{N}{\,\large$2$}
\psfrag{n}{\large$n$}
\center\includegraphics[width=5cm]{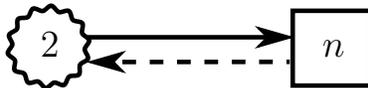}
\caption{\label{fig:fundW2n}Quiver representation of the Wilson loop 
in \eqref{2N|1}, in the same theory as in Figure~\ref{QfundW}.}
\end{centering}
\end{figure}
Going back to the theory with $n$ fundamentals, 
we can construct more complicated Wilson loops based on $3\times3$ block matrices, 
coupling to more than one copy of the gauge field, represented by the 
quiver diagram in Figure~\ref{fig:fundW2n}. 
Let us label the couplings $u_i$, $\bar u^i$, $u'_i$ and $\bar u^{\prime i}$ such that we have 
(note that $G$ and $\bar G$ are nilpotent)
\beq
\label{2N|1}
\cL_0=\begin{pmatrix}A_\mu\frac{\dot x^\mu}{|\dot x|}-i\sigma+\frac{1}{2R}&0&0\\
0&A_\mu\frac{\dot x^\mu}{|\dot x|}-i\sigma+\frac{1}{2R}&0\\
0&0&0\end{pmatrix},
\quad
G=\begin{pmatrix}0&0& u_i\phi^i\\0&0& u'_i\phi^i\\0&0&0\end{pmatrix},
\quad
\bar G=\begin{pmatrix}0&0&0\\0&0&0\\\bar u^i\bar\phi_i&\bar u^{\prime i}\bar\phi_i&0\end{pmatrix}.
\eeq
Now we have the gauge freedom of $GL(2,\bC)$ acting on the $2\times2$ block. Giving the 
moduli space 
$(\bC^{2n})^2\sslash GL(2,\bC)$ (with inverse action on one of the two vector spaces). Again some 
care is needed to identify the singular orbits of this action and reduce to the conical space. 
Starting with the case of a single chiral ($n=1$), the dimension of the quotient seems to vanish and in 
fact there are only singular orbits. The quotient acts on the rank-one matrix constructed out of 
$u\bar u$, $u\bar u'$, $u'\bar u$, $u'\bar u'$ 
by conjugation and can therefore diagonalize it, amounting to setting $u'=\bar u'=0$. 
This implies that the Wilson loops in this case are reducible 
to a block diagonal form with one $2\times2$ block as 
in \eqref{Lopen} and another block with only the gauge field and $\sigma$ and no coupling to the 
chirals, as in \eqref{bos}. The same is true if we include more copies of the gauge field in the 
theory with a single chiral, but for more than one chiral ($n>1$) these spaces are more interesting.

The analysis in the case of $1/4$ BPS Wilson loops is identical, the moduli space can be 
constructed from the quiver describing the Wilson loop.

Taking the example of the theory with $n$ fundamental and $\tilde n$ anti-fundamental chiral fields, 
the discussion after 
\eqref{tilde} suggests two possible constructions of $1/4$ BPS loops. We can get 
$1/2$ BPS loops based on \eqref{1/2,0,-1/2} with say $v=\bar u=0$, so coupling to 
only $\phi$ and $\bar{\tilde\phi}$. In this case, not only $G^2=\bar G^2=0$, but also 
$(G+\bar G)^2=0$, which is similar to the upper-triangular example 
discussed above. These Wilson loops (class III and IV of 
\cite{Ouyang:2015iza,Ouyang:2015bmy,Mauri:2017whf,Mauri:2018fsf}) are 
identical to the analog bosonic Wilson loops. Ignoring such trivial Wilson loops, 
we have a moduli space which is the union of two cones meeting at a point:
$(\bC^{n})^2\sslash\bC^\star_{1,-1}\oplus(\bC^{\tilde n})^2\sslash\bC^\star_{1,-1}$

The moduli space of $1/4$ BPS loops is much larger. The construction 
in \eqref{lat} is similar to \eqref{G} but with a pair of $n+\tilde n$ fields, 
so the moduli space s $(\bC^{n+\tilde n})^2\sslash\bC^\star_{1,-1}$.

The construction in \eqref{1/2,0,-1/2} is similar to \eqref{2N|1}, but the 
symmetry is only $(\bC^*)^2$, so the moduli space is 
$(\bC^{n})^2\sslash\bC^\star_{1,-1}\otimes(\bC^{\tilde n})^2\sslash\bC^\star_{1,-1}$. 
Since this construction anyhow preserves only one supercharge, there 
is really no reason to restrict the blocks to only the $n$ or $\tilde n$ fields, we could 
restore the full effective flavor group and with two copies of matter blocks have 
the moduli space $(\bC^{2n+2\tilde n})^2\sslash GL(2,\bC)$.

\subsection{Quiver varieties}

Analyzing the full spectrum of BPS Wilson loops seems to be a daunting problem where starting 
from the most general $\cL_0$, $\cG$, we need to understand the residual symmetry 
and analyze the resulting quotient.

Luckily this problem is exactly the theory of \emph{quiver varieties} 
(see e.g. \cite{Nakajima1999lectures, Crawley-Boevey, ginzburg2009lectures, kirillov2016quiver})
which is where the full power of the quiver representation of the BPS Wilson loops 
presented in the figures in Section~\ref{sec:WL} comes to the fore. 
To make the connection, recall that a representation of a quiver corresponds 
to assigning a vector space to each node and a linear map to each edge. 
We identified $\cL_0$ with the 
node information, since it includes the gauge fields. In particular, the vector space is $\bC^p$ if there 
are $p$ blocks with the same gauge field (with identical shifts). $G$ and $\bar G$ 
represent the linear maps, as they have entries for each chiral (and anti-chiral) 
field between the appropriate blocks in $\cL_0$. 
Note that $G$ and $\bar G$ have independent $u$ and $\bar u$ 
parameters. As far as representations of quivers, this implies the doubling of edges, where every 
arrow is augmented by another one with the opposite orientation, the dashed arrows 
in our quivers.

We stress that the ranks of the gauge groups play no role here. The overall size of the matrices 
depends on them, but we are not higgsing the vector multiplet, so have no freedom within 
each $N\times N$ block associated to an $SU(N)$ gauge field and it counts as a single copy of 
$\bC$ in the quiver representation. This is different for flavor nodes (or \emph{framing} in the 
mathematical language), where as we saw in the 
examples above, a single $SU(n)$ flavor node introduced a copy of $\bC^n$. Also, in this case we 
do not mod out by the action of $GL(n,\bC)$, as we distinguish between the different 
fundamental fields (say, by assigning them masses).

The spaces of the $u$, $\bar u$ parameters modulo symmetry are exactly the varieties 
associated to the corresponding quiver. 
The classification of quiver representations reproduces all the moduli spaces outlined above 
and provides the answer for the classification of all Wilson loop operators in any other 3d quiver 
gauge theory. This is the main result of this paper and provides an organizing principle to 
previous attempts at classifying such line operators. This also raises many questions, under 
current investigation:

\begin{itemize}
\item
Quiver varieties appear in other contexts in supersymmetric field theories, most 
notably in the question of the Coulomb branch of 3d $\cN=4$ theories, as pioneered 
by Hanany and Witten \cite{Hanany:1996ie}. Why the classification 
of line operators and of Coulomb vacua may be related is unclear.

\item
The spaces found here are all cones, and except for special cases like $\bC$, are 
singular. There are natural resolutions of these spaces, and it would be interesting to 
find whether there is a way to deform the loop operators such that the moduli spaces 
of the deformed loop operators are the resolved spaces. It would also be interesting 
to understand whether more equivalences between loops exist, possibly for specific 
theories, leading to more intricate singular loci.

\item
Though we used the term moduli space, it may be more appropriate to call them 
parameter spaces. It would be interesting to understand whether it has a physical 
interpretation, such that the metric is meaningful.
A natural question is then whether this structure persists at the 
quantum level, as studied for the finite degeneracy of loops preserving 
8 supercharges in \cite{Griguolo:2015swa, Bianchi:2016vvm}. Is the degeneracy 
lifted and/or is the metric corrected. 

\item
Assel and Gomis studied $1/2$ BPS line operators (Wilson and vortex loops) in 
3d $\cN=4$ theories \cite{Assel:2015oxa}. Their operators have double the amount of 
supersymmetry studied here, so it is not surprising that the answers are different. Still, 
it would be good to connect to that work by either specializing the current work to 
enforce more supersymmetry, or generalize their work to less supersymmetric theories.

\item
Related to that, it would be interesting to study the moduli spaces of vortex loops.

\item
It would be interesting to understand the holographic duals of these Wilson loops in theories 
with known holographic duals, like ABJM theory \cite{Aharony:2008ug}. Despite some recent progress 
on that question \cite{Correa:2019rdk}, a full understanding of the bosonic $1/6$ BPS is still lacking.
\end{itemize}

\section*{Acknowledgements}
I am very grateful to all the collaborators on the recent roadmap paper \cite{Drukker:2019bev} who 
stimulated this new examination of old questions. In particular the 
extensive exchanges with M. Probst, D. Trancanelli and M. Tr\'epanier were the seed of this work. 
I am also grateful to A. Hanany, C. Herzog, Y. Lekili, M. Meineri, D. Panov and E. Segal for fruitful conversations.
This work is supported by an STFC grant number  ST/P000258/1.

\appendix
\section{$\cN=2$ theories on the squashed sphere}
\label{app}

This appendix repeats the construction of BPS Wilson loops presented in Section~\ref{sec:WL} 
for theories with matter fields of non-canonical dimensions on the squashed sphere $S^3_b$ and 
fills in some details about the supersymmetry transformations. 

The issue with non-canonical dimensions arise for theories with $\cN\leq2$, which can have 
non-trivial renormalization group flows such that the IR dimensions differ from the canonical ones. 
In such cases it is possible to construct UV theories with arbitrary dimensions such that the 
result of the localization calculation is a function of these dimensions. Using $F$-extremization 
\cite{Jafferis:2010un,Jafferis:2011zi}
allows then to find the correct IR dimensions and plug it into all other calculations in that theory, 
in this case the Wilson loops. We denote the dimensions of the chiral multiplet fields $(\phi,\psi,F)$ 
by $(\Delta,\Delta+1/2,\Delta+1)$.

For the squashed sphere we use the conventions of \cite{Drukker:2012sr} 
(with the replacement $\varphi_1\to\chi$ and $\varphi_2\to\varphi$), 
which are slight modifications of those in \cite{Hama:2011ea} 
(see footnote 12 of \cite{Drukker:2012sr}). In particular the metric on 
the squashed sphere $S^3_b$ is
\beq
ds^2=R^2\left(f(\vartheta)^2d\vartheta^2+b^{2}\sin^2\vartheta\,d\chi^2+b^{-2}\cos^2\vartheta\,d\varphi^2\right),
\qquad
f(\vartheta)=\sqrt{b^{-2}\sin^2\vartheta+b^{2}\cos^2\vartheta}\,.
\eeq
and dreibein
\beq
e^1=Rb^{-1}\cos\vartheta\,d\varphi\,,
\qquad
e^2=-Rb^{}\sin\vartheta\,d\chi\,,
\qquad
e^3=Rf(\vartheta)\,d\vartheta\,.
\eeq

The spinors by default have upper indices are are lowered with $-i\sigma_2$ such that
\beq
\bar\psi\lambda=\lambda\bar\psi\,,
\qquad
\bar\psi\gamma^\mu\lambda=-\lambda\gamma^\mu\bar\psi\,,
\qquad
(\gamma^\mu\bar\psi)\lambda=-\bar\psi\gamma^\mu\lambda\,,
\eeq

The variation of the fields in the vector multiplet are
\bal
\label{delta-vec}
\delta A_\mu&=\tfrac{i}{2}(\bar\epsilon\gamma_\mu\lambda-\bar\lambda\gamma_\mu\epsilon)\,,
\\
\delta\sigma&=\tfrac{1}{2}(\bar\epsilon\lambda-\bar\lambda\epsilon)\,,
\\
\delta\lambda&=-\tfrac{1}{2}\gamma^{\mu\nu}\epsilon F_{\mu\nu}
-D\epsilon+i\gamma^\mu\epsilon D_\mu\sigma
+\tfrac{2i}{3}\sigma\gamma^\mu D_\mu\epsilon\,,
\\
\delta\bar\lambda&=-\tfrac{1}{2}\gamma^{\mu\nu}\bar\epsilon F_{\mu\nu}
+D\bar\epsilon-i\gamma^\mu\bar\epsilon D_\mu\sigma
-\tfrac{2i}{3}\sigma\gamma^\mu D_\mu\bar\epsilon\,,
\\
\delta D
&=-\tfrac{i}{2}\bar\epsilon\gamma^\mu D_\mu\lambda
-\tfrac{i}{2}D_\mu\bar\lambda\gamma^\mu\epsilon
+\tfrac{i}{2}[\bar\lambda\epsilon,\sigma]
+\tfrac{i}{2}[\bar\epsilon\lambda,\sigma]
-\tfrac{i}{6}(D_\mu\bar\epsilon\gamma^\mu\lambda
+\bar\lambda\gamma^\mu D_\mu\epsilon)\,.
\eal
For the chiral multiplet we have
\bal
\label{delta-chi}
\delta\phi&=\bar\epsilon\psi\,,
\qquad
\delta\bar\phi=\epsilon\bar\psi\,,
\\
\delta\psi&=i\gamma^\mu\epsilon D_\mu\phi+i\epsilon\sigma\phi
+\tfrac{2\Delta i}{3}\gamma^\mu D_\mu\epsilon\phi
+\bar\epsilon F\,,
\\
\delta\bar\psi
&=i\gamma^\mu\bar\epsilon D_\mu\bar\phi+i\bar\phi\sigma\bar\epsilon
+\tfrac{2\Delta i}{3}\bar\phi\gamma^\mu D_\mu\bar\epsilon
+\bar F\epsilon\,,
\\
\delta F&=
\epsilon(i\gamma^\mu D_\mu\psi-i\sigma\psi-i\lambda\phi)
+\tfrac{i}{3}(2\Delta-1)D_\mu\epsilon\gamma^\mu\psi\,,
\\
\delta\bar F&=
\bar\epsilon(i\gamma^\mu D_\mu\bar\psi-i\bar\psi\sigma+i\bar\phi\bar\lambda)
+\tfrac{i}{3}(2\Delta-1)D_\mu\bar\epsilon\gamma^\mu\bar\psi\,.
\hskip3.9cm
\eal

For $b\neq1$ a BPS Wilson loop can be either along the $\varphi$ direction at $\vartheta=0$, or 
along $\chi$ at $\vartheta=\pi/2$. We focus on the former, but everything works for the 
other case as well. One also needs to turn on a 
background field that the spinors are charged under
\beq
V_\mu dx^\mu=
-\frac{1}{2}\left(1-\frac{1}{bf(\vartheta)}\right)d\varphi
-\frac{1}{2}\left(1-\frac{b}{f(\vartheta)}\right)d\chi\,.
\eeq

The two supercharges preserving the Wilson loops are parametrized by the Killing spinors
\beq
\qquad
\epsilon=\frac{1}{\sqrt2}
\begin{pmatrix}
e^{i(\varphi+\chi+\vartheta)/2}\\
e^{i(\varphi+\chi-\vartheta)/2}\end{pmatrix},
\qquad
\bar\epsilon=\frac{1}{\sqrt2}
\begin{pmatrix}
-e^{-i(\varphi+\chi-\vartheta)/2}\\
e^{-i(\varphi+\chi+\vartheta)/2}\end{pmatrix}.
\eeq
Clearly at $\vartheta=0$ we have $\sigma_1\epsilon=\epsilon$ and
$\sigma_1\bar\epsilon=-\bar\epsilon$. We find
\beq
\bar\epsilon\epsilon\equiv
\bar\epsilon^\alpha\epsilon_\alpha=1\,,
\qquad
v^\mu=\bar\epsilon\gamma^\mu\epsilon=\left(\frac{b}{R},\frac{1}{Rb},0\right).
\eeq

We denote the corresponding supercharges $Q$ and $\bar Q$, such that 
$Q\Psi=\partial_\epsilon \delta\Psi$, and likewise for $\bar Q$.
With $\cQ_\pm=Q\pm\bar Q$ we have the double variation of the scalars in the chiral 
multiplet
\bal
\label{Q2}
\cQ_+^2\phi&=[\delta_\epsilon,\delta_{\bar\epsilon}]\phi
=iv^\mu(\partial_\mu+iA_\mu)\phi+i\sigma\phi-\Delta\left(\frac{1}{Rf(\vartheta)}+v^\mu V_\mu\right)\phi\,,
\\
\cQ_+^2\bar\phi&=[\delta_\epsilon,\delta_{\bar\epsilon}]\bar\phi
=iv^\mu(\partial_\mu\bar\phi-i\bar\phi A_\mu)-i\bar\phi\sigma+\Delta\left(\frac{1}{Rf(\vartheta)}+v^\mu V_\mu\right)\bar\phi\,.
\eal
Similar expressions exist for the other fields, which is necessary to prove closure of the 
off-shell SUSY algebra, but not for the details of our construction.

At $\vartheta=0$ the last term in \eqref{Q2} is $-\Delta(b+1/b)/2R$. In the 
following we parametrize the deviation from $\Delta=1/2$ and $b=1$ using
\beq
\label{Delta'}
\Delta'=\frac{\Delta}{2}\left(b+\frac{1}{b}\right),
\qquad
q=\exp \pi i(2\Delta'-1)=-\exp \pi i\Delta(b+b^{-1})\,.
\eeq
The original bosonic loop \eqref{bos} can then be written (up to a phase and shift) as
\beq
\label{qW}
qW+1=-\sTr\cP \exp\oint i \cL_0\,|\dot x| \,d\tau\,,
\qquad
\cL_0=\begin{pmatrix}
A_\mu\frac{\dot x^\mu}{|\dot x|}-i\sigma+\frac{\Delta'}{R}&
0\\
0&
0
\end{pmatrix}.
\eeq
Such a structure is required in order to couple the Wilson loop to the matter fields. 
Instead of the usual supertrace in \eqref{L0} (and the $1/2$ BPS Wilson loop \cite{Drukker:2009hy}), 
in this case we need a $q$-deformed sum (a simple rescaling gives 
the form $q^{1/2} W+q^{-1/2}$).

Following the steps in the main text we introduce $\cG$ as in \eqref{delta2G}
\beq
\label{a:delta2G}
\cQ_+^2\cG_{u,\bar u}=i{\mathfrak D}_0\cG_{u,\bar u}
=\frac{i}{R}\partial_\varphi\cG_{u,\bar u}-\left[\cL_0,\cG_{u,\bar u}\right],
\qquad
\cG_{u,\bar u}=\frac{1}{R^{\Delta-1/2}}\begin{pmatrix}0& u_i\phi^i\\\bar u^i\par\phi_j&0\end{pmatrix}.
\eeq
We needed to introduce explicit powers of the the radius $R$ into $\cG$, to give it dimension $1/2$. We are 
assuming that all the fundamental fields $\phi^i$ have the same dimension, otherwise we 
require further modifications to a larger superconnection with different shifts in $\cL_0$ 
as in \eqref{1/2,0,-1/2} and appropriate powers of $R$ in the different blocks of 
$\cG$.

We use the new $\cG_{u,\bar u}$ to define a superconnection which has dimension one and 
from it we get the Wilson loop
\beq
W_{u,\bar u}=\sTr\cP \exp\oint i \cL_{u,\bar u}\,|\dot x| \,d\tau\,,
\qquad
\cL_{u,\bar u}=\cL_0-i\cQ_+\cG_{u,\bar u}
+\cG_{u,\bar u}^2\,.
\eeq
As in \eqref{total-der}, we find that the supersymmetry variation is a total derivative
\beq
\cQ_+\cL_{u,\bar u}
=\mathfrak{D}_{u,\bar u}\cG_{u,\bar u}\,.
\eeq
The argument for the cancelation of the boundary terms from integrating this 
is similar to the argument for the requirement of 
$\sTr$ in the main text, or the argument used in \cite{Drukker:2009hy}. 
As discussed around \eqref{1/4}, we can show that this Wilson loop is also invariant under $\cQ_-$ 
if $G_u^2=\bar G_{\bar u}^2=0$, where these are the chiral and anti-chiral parts of 
$\cG_{u,\bar u}$.

Using the anti-fundamental chiral fields also works as before, but now $\widetilde\cL_0$ is
\beq
\widetilde\cL_0=\begin{pmatrix}
A_\mu\frac{\dot x^\mu}{|\dot x|}-i\sigma-\frac{\Delta'}{R}&
0\\
0&
0
\end{pmatrix}.
\eeq
For generic $\Delta'$, this is not gauge equivalent to $\cL_0$. In fact, even before the deformation 
we find that the analog of \eqref{qW} is
\beq
-\sTr\cP \exp\oint i \widetilde\cL_0\,|\dot x| \,d\tau=q^{-1}W+1\,.
\eeq
This is a different linear combination and once we incorporate the matter fields the two constructions 
are really inequivalent. One can combine fundamental and anti-fundamentals as in 
\eqref{1/2,0,-1/2}, but not the ``latitudes'' of \eqref{lat}. 
The construction of the generic Wilson loop in such a theory is outlined 
in Section~\ref{sec:more}.

\bibliographystyle{utphys2}
\bibliography{refs}

\providecommand{\href}[2]{#2}\begingroup\raggedright\begin{thebibliography}{10}\linespread{1}\selectfont\setlength{\parskip}{0pt}\setlength{\itemsep}{1pt
  plus 0.3ex}

\bibitem{Gaiotto:2007qi}
D.~Gaiotto and X.~Yin, ``{Notes on superconformal Chern-Simons-matter
  theories},'' \href{http://dx.doi.org/10.1088/1126-6708/2007/08/056}{{\em
  JHEP} {\bfseries 08} (2007) 056},
\href{http://arxiv.org/abs/0704.3740}{{\ttfamily arXiv:0704.3740}}.
%%CITATION = ARXIV:0704.3740;%%.

\bibitem{Drukker:2008zx}
N.~Drukker, J.~Plefka, and D.~Young, ``{Wilson loops in 3-dimensional $\cN=6$
  supersymmetric Chern-Simons Theory and their string theory duals},''
  \href{http://dx.doi.org/10.1088/1126-6708/2008/11/019}{{\em JHEP} {\bfseries
  11} (2008) 019},
\href{http://arxiv.org/abs/0809.2787}{{\ttfamily arXiv:0809.2787}}.
%%CITATION = ARXIV:0809.2787;%%.

\bibitem{Chen:2008bp}
B.~Chen and J.-B. Wu, ``{Supersymmetric Wilson loops in ${\cal N}=6$ super
  Chern-Simons-matter theory},''
  \href{http://dx.doi.org/10.1016/j.nuclphysb.2009.09.015}{{\em Nucl. Phys.}
  {\bfseries B825} (2010) 38--51},
\href{http://arxiv.org/abs/0809.2863}{{\ttfamily arXiv:0809.2863}}.
%%CITATION = ARXIV:0809.2863;%%.

\bibitem{Rey:2008bh}
S.-J. Rey, T.~Suyama, and S.~Yamaguchi, ``{Wilson loops in superconformal
  Chern-Simons theory and fundamental strings in anti-de Sitter supergravity
  dual},'' \href{http://dx.doi.org/10.1088/1126-6708/2009/03/127}{{\em JHEP}
  {\bfseries 03} (2009) 127},
\href{http://arxiv.org/abs/0809.3786}{{\ttfamily arXiv:0809.3786}}.
%%CITATION = ARXIV:0809.3786;%%.

\bibitem{Drukker:2009hy}
N.~Drukker and D.~Trancanelli, ``{A supermatrix model for $\cN=6$ super
  Chern-Simons-matter theory},''
  \href{http://dx.doi.org/10.1007/JHEP02(2010)058}{{\em JHEP} {\bfseries 02}
  (2010) 058},
\href{http://arxiv.org/abs/0912.3006}{{\ttfamily arXiv:0912.3006}}.
%%CITATION = ARXIV:0912.3006;%%.

\bibitem{Lee:2010hk}
K.-M. Lee and S.~Lee, ``{$1/2$-BPS Wilson loops and vortices in ABJM model},''
  \href{http://dx.doi.org/10.1007/JHEP09(2010)004}{{\em JHEP} {\bfseries 09}
  (2010) 004},
\href{http://arxiv.org/abs/1006.5589}{{\ttfamily arXiv:1006.5589}}.
%%CITATION = ARXIV:1006.5589;%%.

\bibitem{Ouyang:2015qma}
H.~Ouyang, J.-B. Wu, and J.-j. Zhang, ``{Supersymmetric Wilson loops in $
  \mathcal{N}=4 $ super Chern-Simons-matter theory},''
  \href{http://dx.doi.org/10.1007/JHEP11(2015)213}{{\em JHEP} {\bfseries 11}
  (2015) 213},
\href{http://arxiv.org/abs/1506.06192}{{\ttfamily arXiv:1506.06192}}.
%%CITATION = ARXIV:1506.06192;%%.

\bibitem{Cooke:2015ila}
M.~Cooke, N.~Drukker, and D.~Trancanelli, ``{A profusion of $1/2$ BPS Wilson
  loops in $\mathcal{N}=4$ Chern-Simons-matter theories},''
  \href{http://dx.doi.org/10.1007/JHEP10(2015)140}{{\em JHEP} {\bfseries 10}
  (2015) 140},
\href{http://arxiv.org/abs/1506.07614}{{\ttfamily arXiv:1506.07614}}.
%%CITATION = ARXIV:1506.07614;%%.

\bibitem{Ouyang:2015iza}
H.~Ouyang, J.-B. Wu, and J.-j. Zhang, ``{Novel BPS Wilson loops in
  three-dimensional quiver Chern-Simons-matter theories},''
  \href{http://dx.doi.org/10.1016/j.physletb.2015.12.021}{{\em Phys. Lett.}
  {\bfseries B753} (2016) 215--220},
\href{http://arxiv.org/abs/1510.05475}{{\ttfamily arXiv:1510.05475}}.
%%CITATION = ARXIV:1510.05475;%%.

\bibitem{Ouyang:2015bmy}
H.~Ouyang, J.-B. Wu, and J.-j. Zhang, ``{Construction and classification of
  novel BPS Wilson loops in quiver Chern-Simons-matter theories},''
  \href{http://dx.doi.org/10.1016/j.nuclphysb.2016.07.018}{{\em Nucl. Phys.}
  {\bfseries B910} (2016) 496--527},
\href{http://arxiv.org/abs/1511.02967}{{\ttfamily arXiv:1511.02967}}.
%%CITATION = ARXIV:1511.02967;%%.

\bibitem{Mauri:2017whf}
A.~Mauri, S.~Penati, and J.-j. Zhang, ``{New BPS Wilson loops in $
  \mathcal{N}=4 $ circular quiver Chern-Simons-matter theories},''
  \href{http://dx.doi.org/10.1007/JHEP11(2017)174}{{\em JHEP} {\bfseries 11}
  (2017) 174},
\href{http://arxiv.org/abs/1709.03972}{{\ttfamily arXiv:1709.03972}}.
%%CITATION = ARXIV:1709.03972;%%.

\bibitem{Mauri:2018fsf}
A.~Mauri, H.~Ouyang, S.~Penati, J.-B. Wu, and J.~Zhang, ``{BPS Wilson loops in
  $ \mathcal{N} \geq2$ superconformal Chern-Simons-matter theories},''
  \href{http://dx.doi.org/10.1007/JHEP11(2018)145}{{\em JHEP} {\bfseries 11}
  (2018) 145},
\href{http://arxiv.org/abs/1808.01397}{{\ttfamily arXiv:1808.01397}}.
%%CITATION = ARXIV:1808.01397;%%.

\bibitem{Drukker:2008jm}
N.~Drukker, J.~Gomis, and D.~Young, ``{Vortex loop operators, M2-branes and
  holography},'' \href{http://dx.doi.org/10.1088/1126-6708/2009/03/004}{{\em
  JHEP} {\bfseries 03} (2009) 004},
\href{http://arxiv.org/abs/0810.4344}{{\ttfamily arXiv:0810.4344}}.
%%CITATION = ARXIV:0810.4344;%%.

\bibitem{Kapustin:2012iw}
A.~Kapustin, B.~Willett, and I.~Yaakov, ``{Exact results for supersymmetric
  abelian vortex loops in 2+1 dimensions},''
  \href{http://dx.doi.org/10.1007/JHEP06(2013)099}{{\em JHEP} {\bfseries 06}
  (2013) 099},
\href{http://arxiv.org/abs/1211.2861}{{\ttfamily arXiv:1211.2861}}.
%%CITATION = ARXIV:1211.2861;%%.

\bibitem{Drukker:2012sr}
N.~Drukker, T.~Okuda, and F.~Passerini, ``{Exact results for vortex loop
  operators in 3d supersymmetric theories},''
  \href{http://dx.doi.org/10.1007/JHEP07(2014)137}{{\em JHEP} {\bfseries 07}
  (2014) 137},
\href{http://arxiv.org/abs/1211.3409}{{\ttfamily arXiv:1211.3409}}.
%%CITATION = ARXIV:1211.3409;%%.

\bibitem{Drukker:2019bev}
N.~Drukker {\em et~al.}, ``{Roadmap on Wilson loops in 3d Chern-Simons-matter
  theories},'' \href{http://dx.doi.org/10.1088/1751-8121/ab5d50}{{\em J. Phys.
  A} {\bfseries 53} no.~17, (2020) 173001},
  \href{http://arxiv.org/abs/1910.00588}{{\ttfamily arXiv:1910.00588}}.

\bibitem{Kapustin:2009kz}
A.~Kapustin, B.~Willett, and I.~Yaakov, ``{Exact results for Wilson loops in
  superconformal Chern-Simons theories with matter},''
  \href{http://dx.doi.org/10.1007/JHEP03(2010)089}{{\em JHEP} {\bfseries 03}
  (2010) 089},
\href{http://arxiv.org/abs/0909.4559}{{\ttfamily arXiv:0909.4559}}.
%%CITATION = ARXIV:0909.4559;%%.

\bibitem{Hama:2010av}
N.~Hama, K.~Hosomichi, and S.~Lee, ``{Notes on SUSY gauge theories on
  three-sphere},'' \href{http://dx.doi.org/10.1007/JHEP03(2011)127}{{\em JHEP}
  {\bfseries 03} (2011) 127},
\href{http://arxiv.org/abs/1012.3512}{{\ttfamily arXiv:1012.3512}}.
%%CITATION = ARXIV:1012.3512;%%.

\bibitem{Hama:2011ea}
N.~Hama, K.~Hosomichi, and S.~Lee, ``{SUSY gauge theories on squashed
  three-spheres},'' \href{http://dx.doi.org/10.1007/JHEP05(2011)014}{{\em JHEP}
  {\bfseries 05} (2011) 014},
\href{http://arxiv.org/abs/1102.4716}{{\ttfamily arXiv:1102.4716}}.
%%CITATION = ARXIV:1102.4716;%%.

\bibitem{Marino:2011nm}
M.~Mari\~no, ``{Lectures on localization and matrix models in supersymmetric
  Chern-Simons-matter theories},''
  \href{http://dx.doi.org/10.1088/1751-8113/44/46/463001}{{\em J. Phys.}
  {\bfseries A44} (2011) 463001},
\href{http://arxiv.org/abs/1104.0783}{{\ttfamily arXiv:1104.0783}}.
%%CITATION = ARXIV:1104.0783;%%.

\bibitem{Cardinali:2012ru}
V.~Cardinali, L.~Griguolo, G.~Martelloni, and D.~Seminara, ``{New
  supersymmetric Wilson loops in ABJ(M) theories},''
  \href{http://dx.doi.org/10.1016/j.physletb.2012.10.051}{{\em Phys. Lett.}
  {\bfseries B718} (2012) 615--619},
\href{http://arxiv.org/abs/1209.4032}{{\ttfamily arXiv:1209.4032}}.
%%CITATION = ARXIV:1209.4032;%%.

\bibitem{Bianchi:2018bke}
M.~S. Bianchi, L.~Griguolo, A.~Mauri, S.~Penati, and D.~Seminara, ``{A matrix
  model for the latitude Wilson loop in ABJM theory},''
  \href{http://dx.doi.org/10.1007/JHEP08(2018)060}{{\em JHEP} {\bfseries 08}
  (2018) 060},
\href{http://arxiv.org/abs/1802.07742}{{\ttfamily arXiv:1802.07742}}.
%%CITATION = ARXIV:1802.07742;%%.

\bibitem{Drukker:2006ga}
N.~Drukker, ``{$1/4$ BPS circular loops, unstable world-sheet instantons and
  the matrix model},''
  \href{http://dx.doi.org/10.1088/1126-6708/2006/09/004}{{\em JHEP} {\bfseries
  09} (2006) 004},
\href{http://arxiv.org/abs/hep-th/0605151}{{\ttfamily hep-th/0605151}}.
%%CITATION = HEP-TH/0605151;%%.

\bibitem{Nakajima1999lectures}
H.~Nakajima, \href{http://dx.doi.org/10.1090/ulect/018}{{\em Lectures on
  Hilbert Schemes of Points on Surfaces}}.
\newblock University Lecture Series. American Mathematical Society, 1999.

\bibitem{Crawley-Boevey}
W.~Crawley-Boevey, ``Lectures on representations of quivers.''
\newblock \url{http://www.math.uni-bielefeld.de/~wcrawley/quivlecs.pdf}.

\bibitem{ginzburg2009lectures}
V.~Ginzburg, ``Lectures on nakajima's quiver varieties,''
  \href{http://arxiv.org/abs/0905.0686}{{\ttfamily arXiv:0905.0686}}.

\bibitem{kirillov2016quiver}
A.~Kirillov, \href{http://dx.doi.org/10.1090/gsm/174}{{\em Quiver
  Representations and Quiver Varieties}}.
\newblock Graduate studies in mathematics. American Mathematical Society, 2016.

\bibitem{Hanany:1996ie}
A.~Hanany and E.~Witten, ``{Type IIB superstrings, BPS monopoles, and
  three-dimensional gauge dynamics},''
  \href{http://dx.doi.org/10.1016/S0550-3213(97)00157-0,
  10.1016/S0550-3213(97)80030-2}{{\em Nucl. Phys.} {\bfseries B492} (1997)
  152--190},
\href{http://arxiv.org/abs/hep-th/9611230}{{\ttfamily hep-th/9611230}}.
%%CITATION = HEP-TH/9611230;%%.

\bibitem{Griguolo:2015swa}
L.~Griguolo, M.~Leoni, A.~Mauri, S.~Penati, and D.~Seminara, ``{Probing Wilson
  loops in ${\cal N}=4$ Chern-Simons-matter theories at weak coupling},''
  \href{http://dx.doi.org/10.1016/j.physletb.2015.12.018}{{\em Phys. Lett. B}
  {\bfseries 753} (2016) 500--505},
  \href{http://arxiv.org/abs/1510.08438}{{\ttfamily arXiv:1510.08438}}.

\bibitem{Bianchi:2016vvm}
M.~S. Bianchi, L.~Griguolo, M.~Leoni, A.~Mauri, S.~Penati, and D.~Seminara,
  ``{The quantum 1/2 BPS Wilson loop in ${\cal N}=4$ Chern-Simons-matter
  theories},'' \href{http://dx.doi.org/10.1007/JHEP09(2016)009}{{\em JHEP}
  {\bfseries 09} (2016) 009}, \href{http://arxiv.org/abs/1606.07058}{{\ttfamily
  arXiv:1606.07058}}.

\bibitem{Assel:2015oxa}
B.~Assel and J.~Gomis, ``{Mirror symmetry and loop operators},''
  \href{http://dx.doi.org/10.1007/JHEP11(2015)055}{{\em JHEP} {\bfseries 11}
  (2015) 055},
\href{http://arxiv.org/abs/1506.01718}{{\ttfamily arXiv:1506.01718}}.
%%CITATION = ARXIV:1506.01718;%%.

\bibitem{Aharony:2008ug}
O.~Aharony, O.~Bergman, D.~L. Jafferis, and J.~Maldacena, ``{${\cal N}=6$
  superconformal Chern-Simons-matter theories, M2-branes and their gravity
  duals},'' \href{http://dx.doi.org/10.1088/1126-6708/2008/10/091}{{\em JHEP}
  {\bfseries 10} (2008) 091}, \href{http://arxiv.org/abs/0806.1218}{{\ttfamily
  arXiv:0806.1218}}.

\bibitem{Correa:2019rdk}
D.~H. Correa, V.~I. Giraldo-Rivera, and G.~A. Silva, ``{Supersymmetric mixed
  boundary conditions in $AdS_{2}$ and DCFT$_{1}$ marginal deformations},''
  \href{http://dx.doi.org/10.1007/JHEP03(2020)010}{{\em JHEP} {\bfseries 03}
  (2020) 010}, \href{http://arxiv.org/abs/1910.04225}{{\ttfamily
  arXiv:1910.04225}}.

\bibitem{Jafferis:2010un}
D.~L. Jafferis, ``{The exact superconformal $R$-symmetry extremizes $Z$},''
  \href{http://dx.doi.org/10.1007/JHEP05(2012)159}{{\em JHEP} {\bfseries 05}
  (2012) 159}, \href{http://arxiv.org/abs/1012.3210}{{\ttfamily
  arXiv:1012.3210}}.

\bibitem{Jafferis:2011zi}
D.~L. Jafferis, I.~R. Klebanov, S.~S. Pufu, and B.~R. Safdi, ``{Towards the
  $F$-theorem: ${\cal N}=2$ field theories on the three-sphere},''
  \href{http://dx.doi.org/10.1007/JHEP06(2011)102}{{\em JHEP} {\bfseries 06}
  (2011) 102}, \href{http://arxiv.org/abs/1103.1181}{{\ttfamily
  arXiv:1103.1181}}.

\end{thebibliography}\endgroup

\end{document}